\documentstyle[floats,aps,prc]{revtex}

\def\be{\begin{equation}}
\def\ee{\end{equation}}
\def\ba{\begin{eqnarray}}
\def\ea{\end{eqnarray}}
\def\bsa{\begin{subeqnarray}}
\def\esa{\end{subeqnarray}}

\begin{document}
\draft
\title{
Bremsstrahlung in $\alpha$ Decay} 
\author{N. Takigawa,$^{1}$ Y. Nozawa,$^{1}$ K. Hagino, $^{1,2}$ 
A. Ono, $^{1}$ and D.M. Brink $^{3}$}
\address{
$^{1}$Department of Physics,
Tohoku University, Sendai 980--8578, Japan \\
$^{2}$Institute for Nuclear Theory, University of Washington, Seattle, 
Washington 98915 \\
$^{3}$Universit\`a degli Studi di Trento
and ECT$^{\large *}$, 
Villa Tambosi, I--38050 Villazzano, 
Trento, Italy
}

\maketitle

\begin{abstract}
A quantum mechanical analysis of the bremsstrahlung in $\alpha$ 
decay of $^{210}$Po is performed in close reference 
to a semiclassical theory. 
We clarify the contribution from the 
tunneling, mixed, outside barrier regions and from the wall 
of the inner potential well to the final 
spectral distribution, and discuss their interplay. 
We also comment on  
the validity of semiclassical calculations, and the possibility 
to eliminate the ambiguity in the nuclear potential 
between the alpha particle and daughter nucleus using 
the bremsstrahlung spectrum. 
\end{abstract}



Bremsstrahlung during $\alpha $ decay and fission is one of 
the intriguing current topics of nuclear physics. 
Kasagi et al. measured the bremsstrahlung in the 
$\alpha $ decay from $^{210}$Po, and 
showed that the emission of low energy photons is 
systematically hindered from what one expects from the pure 
Coulomb acceleration of the $\alpha $ particle in the  
outside region of the potential barrier. They speculated that 
this reduction is caused by a destructive interference between the 
radiations in the tunneling and classically allowed regions, and 
thus might offer a
possibility to learn about the tunneling time. 
They also calculated the spectral distribution 
based on a semiclassical theory of Dyakonov and 
Gornyi \cite{Dyakonov}. Their calculations 
reproduce the experimental cross section well for low energy 
photons. In addition, an interesting thing is that 
their calculations give a hump in agreement with the measurement, 
which however has large error bars in that region.  
On the other hand, the intensity of the radiation 
observed by D'Arrigo et al. \cite{D'Arrigo} for the alpha decay from 
$^{226}$Ra and $^{214}$Po 
is larger than the prediction of the pure Coulomb acceleration model.
Also, their data show 
no structure in the spectral 
distribution as expected from a classical formula. 
In nuclear fission, 
a measurement by Luke et al. gave 
an upper bound to the bremsstrahlung rate for 
the spontaneous fission of $^{252}$Cf.

Recently, Papenbrock and Bertsch performed a quantum mechanical 
calculation for the bremsstrahlung in $\alpha $ decay 
in perturbation theory \cite{PB}. 
Their spectral distribution is 
monotonic, though it is within the error bars of the data in 
Ref.\cite{Kasagi}. The authors claim that the contribution 
from the tunneling wave function under the barrier is small. 
However, the definition of either classical or tunneling is not unique. 
More studies from various points of view are clearly needed 
in order to clarify the situation. 
The aim of this paper is to resolve these puzzling problems 
relating to the spectral distribution, especially on the role 
of the tunneling region. To this end, we perform a quantum mechanical 
analysis in a way which has a close relation to semiclassical theory. 
As we show later, quantum mechanical calculations are required to have 
quantitative accuracy, while a semiclassical theory 
provides a clear physical 
understanding of the phenomena by bridging the quantum mechanical and 
classical calculations of the bremsstrahlung. It also 
provides a more definitive understanding of the 
role of quantum tunneling by naturally dividing the whole region into  
the classical, mixed and tunneling regions through classical 
turning points.  

Fermi's golden rule gives the following expression for the 
photon emission per photon energy during $\alpha $ decay 
in the dipole approximation, 
\be
{dP\over d E_\gamma} = \frac{4 Z_{\rm \tiny eff}^2e^2}{3 m^2 c^3}
  \left|\langle\Phi_f\left|\partial_r V\right|\Phi_i\rangle\right|^2
  {1\over E_\gamma},
\label{fermi}
\ee
where $V(r)$ is the interaction between the $\alpha $ particle 
and the daughter nucleus 
as a function of their separation distance $r$, 
$m$ is their reduced mass and 
$E_\gamma $ is the photon energy. The effective charge $Z_{\rm \tiny eff}$ 
is given by $Z_{\rm \tiny eff}=(A_DZ_{\alpha}-A_{\alpha}Z_D)/A_P$, where 
$A_D,Z_D,A_{\alpha},Z_{\alpha}$  and $A_P$ 
are the mass and atomic numbers of the daughter 
nucleus and of $\alpha $ particle, and the mass number of the parent nucleus, 
respectively. $\Phi_i(r)$ and $\Phi_f(r)$ 
are the appropriately normalized radial wave functions of the 
initial and final states of the $\alpha $ particle, respectively\cite{PB}. 

As in Ref.\cite{PB}, we take the following simple model 
for the interaction 
between the $\alpha $ particle and the daughter nucleus 
\be
V(r) = (Z_{\alpha}Z_De^2/r)\Theta(r-r_0) - V_0\Theta(r_0-r). 
\label{pot}
\ee
The initial and final state wave functions are given by
\ba
\Phi_i &=& \left\{
\begin{array}{ll}
({m/\hbar k_{i}})^{1\over 2}     
 R_i(r)/r & (r>r_0) \\
N_i j_0(K_i r) &(r<r_0)
\end{array}\right.
\label{phiii}
\\
\Phi_f &=& \left\{
\begin{array}{ll}
(2m/\pi\hbar^2 k_{f})^{1\over 2}
R_f(r)/r & (r>r_0)\\
N_f j_1(K_f r) &(r<r_0)
\end{array}\right.
\label{phifi}
\ea
where $R_i(r)=G_0(\eta_i,k_{i}r)+iF_0(\eta_i,k_{i}r)$ and
$R_f(r)=F_1(\eta_f,k_{f}r)\cos\alpha+G_1(\eta_f,k_{f}r)\sin\alpha$, 
$\alpha$ being the phase shift.
The wave numbers in these equations are given by,
\ba
K_i &=& \hbar^{-1}\sqrt{2m(E_i+V_0)};\quad k_{i}=
\hbar^{-1}\sqrt{2mE_i}\\    
K_f &=& \hbar^{-1}\sqrt{2m(E_f+V_0)};\quad k_{f}=
\hbar^{-1}\sqrt{2mE_f}      
\ea
where $E_i=E_\alpha$, $E_\alpha $ being the energy of the $\alpha $ decay,  
$E_f=E_\alpha - E_\gamma=E_\alpha - \hbar \omega$ 
and $\eta$ the Sommerfeld parameter.

The transition matrix in Eq.(\ref{fermi}) now reads, 
\ba
&&\langle\Phi_f\left|\partial_r V\right|\Phi_i\rangle \nonumber\\
&&\qquad = \sqrt{2m^2\over\pi\hbar^3k_{i}k_{f}}\bigg\{I_W
+\int_{r_0}^\infty dr\,A(r) R_f(r)R_i(r)\bigg\},
\label{me}
\ea
where $A(r)$ stands for  
$\partial_r V=-Z_{\alpha}Z_De^2r^{-2}$, and 
 the wall contribution $I_W$ is given by
\be
I_W=(Z_{\alpha}Z_De^2/r_0+V_0)
R_f(r_0)R_i(r_0) 
\label{wall}
\ee
\noindent

If we denote the external turning points in the 
initial and final states by $r_{ei}$ and $r_{ef}$, 
respectively, the integration in Eq.(\ref{me}) can be naturally 
divided into three parts, i.e. 
the integrations between $r_0$ and $r_{ei}$, between $r_{ei}$ and $r_{ef}$ 
and between $r_{ef}$ and $\infty$. We call them 
the tunneling, mixed and outside regions, respectively. 
Straight-forward calculations of each integral using the exact Coulomb 
wave functions will clarify the role of each term, especially the 
role of the tunneling region. Before we present the results of 
such quantum mechanical calculations, we wish to 
present semiclassical formulae for radiation. 

They can be derived by replacing the quantum mechanical 
Coulomb wave functions by 
their semiclassical representations, 
which read \cite{brink}
\ba
F_\ell &\sim&\left\{
\begin{array}{ll}
 (\frac{k}{k_\ell(r)})^{\frac{1}{2}} \sin (\int^r_{r_e} 
k_\ell(r) dr + \frac{\pi}{4} ) & \mbox{($r>r_e$)} \\
 \frac{1}{2}(\frac{k}{\kappa_\ell(r)})^{\frac{1}{2}} 
\exp(-\int^{r_e}_r \kappa_\ell(r) dr ) &\mbox{($r<r_e$)}
\end{array}\right. 
\label{cwff}
\\
G_\ell &\sim& \left\{
\begin{array}{ll}
 (\frac{k}{k_\ell(r)})^{\frac{1}{2}} \cos (\int^r_{r_e} 
k_\ell(r) dr + \frac{\pi}{4} ) & \mbox{($r>r_e$)} 
\\
  (\frac{k}{\kappa_\ell(r)})^{\frac{1}{2}} 
\exp(\int^{r_e}_r \kappa_\ell(r) dr ) & \mbox{($r<r_e$)}
\end{array}\right.
\label{cwfg}
\ea
where $r_e$ is the external classical turning point, and $k_\ell $ and
$\kappa_\ell$ are the wave numbers for the orbital angular momentum
$\ell$ in the classically allowed and forbidden regions, respectively.
Notice that $G_\ell$ and $F_\ell$ increases and decreases
exponentially, respectively, as $r$ moves deeper inside the tunneling
region. We remark that the phase shift $\alpha $ in Eq.(\ref{phifi})
is proportional to the tunneling probability and is very small. This
can be proved by matching the semiclassical wave function under the
potential barrier to the wave function in the region of the potential
well.  We thus express $\tan \alpha$ as
\be
\tan \alpha=C_\alpha P_{tf}, 
\qquad
P_{tf}=e^{-2\int^{r_{ef}}_{r_0} \kappa_{1f}(r)dr},
\label{phasea}
\ee
where $P_{tf}$ is the tunneling probability in the final state in the
WKB approximation. In Eq.(\ref{phasea}) and in what follows, the meaning
of the lower indices will be obvious, e.g. $\kappa_{1f}$
is the wave number with angular
momentum 1 in the final state.

We now derive the semiclassical formulae for the contribution to 
the integral in Eq.(\ref{me}) from the tunneling, mixed and outside 
regions separately.

\noindent 
{\underline{1. Tunneling region:}}
The main contribution from the tunneling region comes from the 
$F_1\cdot G_0$ term, whose semiclassical representation is given by
\ifpreprintsty\begin{footnotesize}\fi
\widetext
\ba
\int^{r_{ei}}_{r_0} A(r) F_1(\eta_f,k_fr) \ G_0(\eta_i,k_ir) dr
\sim
\frac{1}{2}\sqrt{k_{i}k_{f}}
e^{-\int^{r_{ef}}_{r_{ei}}\kappa_{1f}(r) dr}
\int^{r_{ei}}_{r_0} A(r) \biggl(\frac{1}{\kappa_{0i}(r)\kappa_{1f}(r)}\biggr)^{\frac12}
e^{-\int^{r_{ei}}_r(\kappa_{1f}(r')-\kappa_{0i}(r')) dr'} dr
 \\
\sim
\frac{\hbar\sqrt{k_{i}k_{f}}}{2m}
e^{-\int^{r_{ef}}_{r_{ei}}\kappa_{1f}(r) dr}
\int^{T_i}_0 A(r(\tau_i)) \biggl(\frac{\kappa_{0i}(r(\tau_i))}{\kappa_{1f}
(r(\tau_i))}\biggr)^{\frac12}
e^{-\frac{1}{\hbar}\int^{T_i}_{\tau_i} (\frac{\hbar^21(1+1)}{2m(r(\tau'_i))^2}+\hbar \omega) d\tau'_i}d\tau_i\label{intt}
\ea
\noindent
\ifpreprintsty\end{footnotesize}\fi
In obtaining the last expression, we expanded the  
$\kappa_{1f}(r)-\kappa_{0i}(r)$ up to the leading order of $\frac{\hbar^21(1+1)}{2m(r(\tau'))^2}+\hbar \omega $ and introduced the 
time parameter along the (imaginary) time axis 
$\tau_i $ by 
\be
\tau_i(r)=\int^r_{r_0}\frac{dr}{v_i(r)},\quad
v_i(r)=\sqrt{\frac{2}{m}\left(\frac{Z_{\alpha}Z_De^2}{r}-E_i\right)}
\ee
$T_i=\tau_i(r_{ei})$ is the tunneling time in the initial channel. 
 The lower index $i$ of $\tau_i$ shows that the time
is related to the distance $r$ through the velocity
 in the initial state. Notice that the radiation amplitude in
this region is given by the Laplace transform of the acceleration if 
one could ignore the centrifugal term.
The $G_1\cdot G_0$ term is the largest as long as the
integral itself is concerned.
However, the contribution of this term is reduced by the small value
of $\tan\alpha$ in front of $G_1$.
 We have a similar expression
for the $G_1\cdot F_0$ term, which is also expected to be small. The
$F_1\cdot F_0$ term is expected to be the smallest.

\noindent
{\underline{2. Mixed region:}} 
The mixed region, which is classically allowed in the initial 
state but forbidden after radiation, is beyond the scope of classical 
calculations in Refs.\cite{Kasagi} and \cite{Luke}. 
Although quantum mechanical calculations 
are required to quantitatively estimate the contribution from this region 
as we show later, 
it is still interesting to see its semiclassical representation. 
The contribution of the $F_1$ term which dominates in this region reads, 
\widetext
\be
\int^{r_{ef}}_{r_{ei}} A(r) \ F_1(\eta_f,k_fr)\ R_i(r) dr
\sim \frac{1}{2}
\int^{r_{ef}}_{r_{ei}} A(r) 
\biggl(\frac{k_{i}k_{f}}{k_{0i}(r)\kappa_{1f}(r)}\biggr)^{\frac12}
e^{i\bigl[\int^r_{r_{ei}}k_{0i}(r') dr'+\frac{\pi}{4} \bigr] }
e^{-\int^{r_{ef}}_r\kappa_{1f}(r') dr'} dr
\ee
\noindent
\noindent
We show later that 
this region strongly influences the behaviour of the 
spectral distribution at high energies.

\noindent
{\underline{3. Outside the barrier:}}  
The main contribution in the classical 
region, i.e. outside the barrier, comes from the $F_1$
term. Its semiclassical expression reads, 
\ifpreprintsty\begin{footnotesize}\fi
\widetext
\be
\int^{\infty}_{r_{ef}} A(r)\ F_1(\eta_f,k_fr)\ R_i(r) dr 
\sim -\frac{\hbar\sqrt{k_{i}k_{f}}}{2im}
e^{i\int^{r_{ef}}_{r_{ei}} k_{0i}(r) dr}
\int^{\infty}_0 A(r(t_f)) 
\biggl(\frac{k_{1f}(r(t_f))}{k_{0i}(r(t_f))}\biggr)^{\frac12}
e^{\frac{i}{\hbar}\int^{t_f}_0 (\frac{\hbar^21(1+1)}{2m(r(t'_f))^2}+\hbar \omega)
 dt'_f} dt_f
\ee
\noindent
\ifpreprintsty\end{footnotesize}\fi
This is nothing but the classical formula for the bremsstrahlung except 
for the existence of the centrifugal potential term.  The
lower index $f$ of $t_f$ means that the time $t$ is
related to the distance $r$ through the velocity in the final
state. We have ignored the term where the action integrals in the
initial and final states enter with the same sign, because the sign of
the integrand in that case rapidly changes as a function of $r$
and the value of integration becomes negligibly small.

We next consider the contribution from the wall. 
This term has been overlooked in 
the calculations reported in Ref.\cite{Kasagi}. 
As we clearly demonstrate later, this is one of the key issues to resolve the 
discrepancy between the calculations in Refs.\cite{Kasagi} and 
\cite{PB}. The contribution of the $F_0$ term can be safely ignored 
unless the energy of the $\alpha$ particle is very close to the top of
the potential barrier. 
Using Eq.(\ref{phasea}), the remaining terms can be expressed as,
\widetext
\be
R_f(r_0)G_0(\eta_i,k_{i}r_0)
\sim\frac{1}{2}\biggl(\frac{k_{f}k_{i}}
{\kappa_{1f}(r_0)\kappa_{0i}(r_0)}\biggr)^{\frac12}
\cos\alpha(1+2C_{\alpha})e^{-\int^{r_{ef}}_{r_{ei}}\kappa_{1f}(r) dr}
e^{-\frac{1}{\hbar}\int^{T_i}_{0} (\frac{\hbar^21(1+1)}{2m(r(\tau_i))^2}
+\hbar \omega) d\tau_i}.
\label{iwgb}
\ee
\noindent

We now analyse the bremsstrahlung in the $\alpha $
decay from $^{210}$Po, where $E_{\alpha}$=5.3 MeV. 
We determine the potential parameters in the same way as in Ref.\cite{PB}, 
i.e. by matching the quantum mechanical wave functions at $r_0$, 
and by calculating the decay width 
by normalizing the current at $r_e$ to the probability 
inside. 
An alternative is to use the semiclassical 
quantization rule and the 
R-matrix theory for the decay rate (see Eqs.(3) and (4) in Ref.\cite{Buck}).
We found, however, that the semi-classical quantization rule 
is not valid with 
significant influences on the bremsstrahlung spectrum (see later). 

Figure 1 shows the spectral distribution calculated quantum mechanically
with one of the potential parameter sets, $V_0$ = 21.37 MeV and $r_0$
= 8.055 fm, which gives 5 nodes to the wave function for the radial
motion. The general behaviour of the theoretical results do not
depend so much on the particular choice of the potential parameters
(see below, however).  This potential parameter set corresponds to
that in Refs.\cite{PB,PBpot}.  The thick solid line represents the
total photon emission probability.  It is a monotonically decreasing
function and is consistent with the data except for the last data
point.  The figure also shows that various components have nearly the
same order of magnitude, and that the final spectrum is the result of
a complicated interference among them. In order to see the role of
tunneling and classical regions and the interplay of each contribution
more clearly, we show in Fig.2 specral distributions by dividing into
the classical and tunneling contributions. We present two different
combinations depending on whether we consider the mixed region
classical or tunneling. We call the part containing the outside
barrier component the classical.  In either case, the soft photon
emission is dominated by classical contributions. This follows from
the strong cancellation between the wall contribution and the
contribution from the integral in the tunneling region.  If the
potential has a finite slope at the inner turning point and if one
could ignore the centrifugal term, the net contribution from the
tunneling region should be zero in the soft photon limit. Our result
matches with this general aspect.  For high energy photons, the
classical and the tunneling contributions have nearly the same
magnitude and interfere destructively leading to a much smaller photon
emission than the the prediction of the classical theory. The
dot-dashed line in Fig.2 shows the photon emission probability
obtained by summing the contributions from the outside barrier and
tunneling regions. Interestingly, this has a hump at high photon
energies, just like that discussed in Ref.\cite{Kasagi}, which
overlooked the wall contribution and ignored the contribution from the
mixed region. However, this interesting interference pattern is
hidden by these contributions.

As well known, the resonance position and the decay width do not
uniquely determine the potential. Each potential gives a different
value for the coefficient $C_\alpha$ in our formalism.  An interesting
question is then whether the bremsstrahlung can be used to give
additional constraints to the potential. If one restricts to the
simple square well potential in Eq.(\ref{pot}), various sets
correspond to different number of nodes $\it n$ in the radial wave
function inside the potential well. We found that the spectral
distribution is almost indistinguishable as long as ${\it n} \leq
7$. For larger values of ${\it n}$, one starts to notice a difference
in that the soft photon emission probability is predicted to be
recognizably larger than the prediction of the classical theory. 
Note that the soft photons mean here those whose energy is, say, 
smaller than  100 keV. The emission probability of zero energy 
photons is not influenced by the potential parameters as discussed in 
Ref.\cite{PB}.
As an example, we show in Fig.3 the spectral distribution calculated by
the potential set, $V_0$ = 107.3 MeV and $r_0$ = 7.947 fm, which
corresponds to ${\it n}=11$ and matches with a simple estimate based
on the Pauli principle.  In this respect, it would be extremely
interesting to extend the data towards lower photon energies to
examine whether the shallow or the deep potential agrees better 
with the data of bremsstrahlung. $C_\alpha $ gradually decreases 
from 0.041 to $-$0.055, and from $-$0.25 to $-$0.28 
 for the potentials in Figs.1 and 3, respectively, 
as $E_\gamma$ changes from 0 to 600 keV.

We finally comment on the validity of semiclassical calculations. We
already mentioned a problem with the semiclassical quantization rule
in determining the potential parameters. Since 
$C_\alpha$ is sensitive to the slope of the wave function at the
matching point, it causes a serious error in properly describing the
bremsstarhlung spectrum. We found that the wall contribution itself
gets very small because of the cancellation between the $F_1$ and
$G_1$ terms in Eq.(\ref{iwgb}) if we adopt any potential set
determined by using the semiclassical quantization rule. Consequently,
the contribution from the tunneling region dominates the spectrum at
soft photons giving larger cross section than the classical
prediction. Another problem is that two turning points $r_{ei}$ and
$r_{ef}$ lie too close to each other to use naive semiclassical wave
functions. The crosses in Fig.1 represent the contribution from the
mixed region calculated semiclassically. The deviation from the dotted
line clearly shows the failure of the semiclassical calculations.
Since the contribution from the mixed region strongly influences high
energy spectrum, this is a serious problem.  The semiclassical
calculations reproduce the qualitative behaviour of the quantum
mechanical results quite well 
for the contributions from the regions under and outside the barrier,
but have some quantitative inaccuracy especially at high energies.

In summary, our analysis shows that the final bremsstrahlung spectrum 
results from a subtle interferences of the contributions of 
the tunneling, mixed and classical regions as well as the wall of the 
the potential well, each 
of which has a comparable magnitude.
Semiclassical as well as classical theories seem not to be reliable 
for describing these subtle interference effects, though they 
give some clear understanding of the phenomena. 
It will be very interesting if one could perform more exclusive 
experiments to pick up each contribution separately. 
The extention of data to more soft photon side 
is awaited to provide a stringent test of the potential between the 
alpha particle and the daughter nucleus.
Our study is based on the assumption of the validity 
of the potential model. Though this assumption seems to work in our 
analysis, it is a very interesting question to examine 
more in detail whether the bremsstrahlung spectrum contains 
some information beyond the scope of the  potential model 
such as the validity of the R-matrix theory and 
the preformation factor.

\bigskip

We thank G.F. Bertsch, T. Papenbrock, A. Bulgac, M. Abe and J. Kasagi 
for discussions. 
This research was supported by
the Monbusho International Scientific Research
Program: Joint Research: contract number 09044051;
and by Grant-in-Aid for General Scientific
Research, Contract number 08640380 from the Japanese
Ministry of Education, Science and Culture.
The work of K.H. was supported by the Japan Society for the Promotion of
Science for Young Scientists.

\newpage

\noindent
{\bf Figure Captions:}

\bigskip

\noindent
{\bf Fig. 1:}Total photon emission probability and its decomposition.
The potential parameters are $V_0=21.37$ MeV and $r_0=8.055$ fm.
The crosses are the mixed region contribution calculated semiclassically. 
All the other lines have been obtained quantum mechanically. 
The data are from Ref. \protect\cite{Kasagi}.

\medskip

\noindent
{\bf Fig. 2:}
Grouping of each component into classical and 
tunneling contributions. The potential parameters are 
the same as those for Fig. 1.

\medskip

\noindent
{\bf Fig. 3:}The same as Fig.2, but for $V_0=107.3$ MeV and $r_0=7.947$ fm.


\begin{thebibliography}{99}
\bibitem{PBpot}
We found a mistake 
in the parameter sets in Ref.\protect{\cite{PB}}. However, the physics  
does not change so much.
\end{thebibliography}
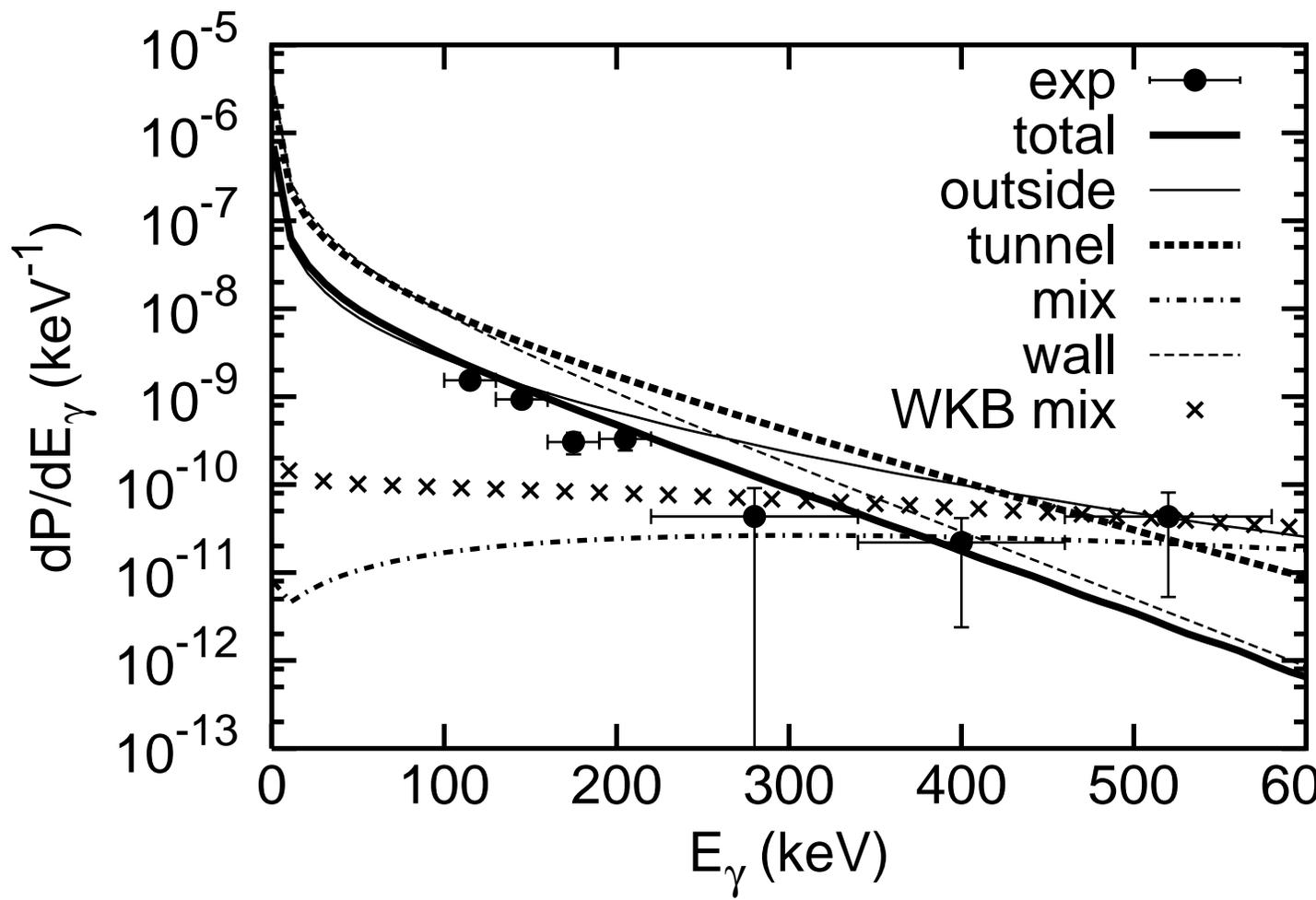
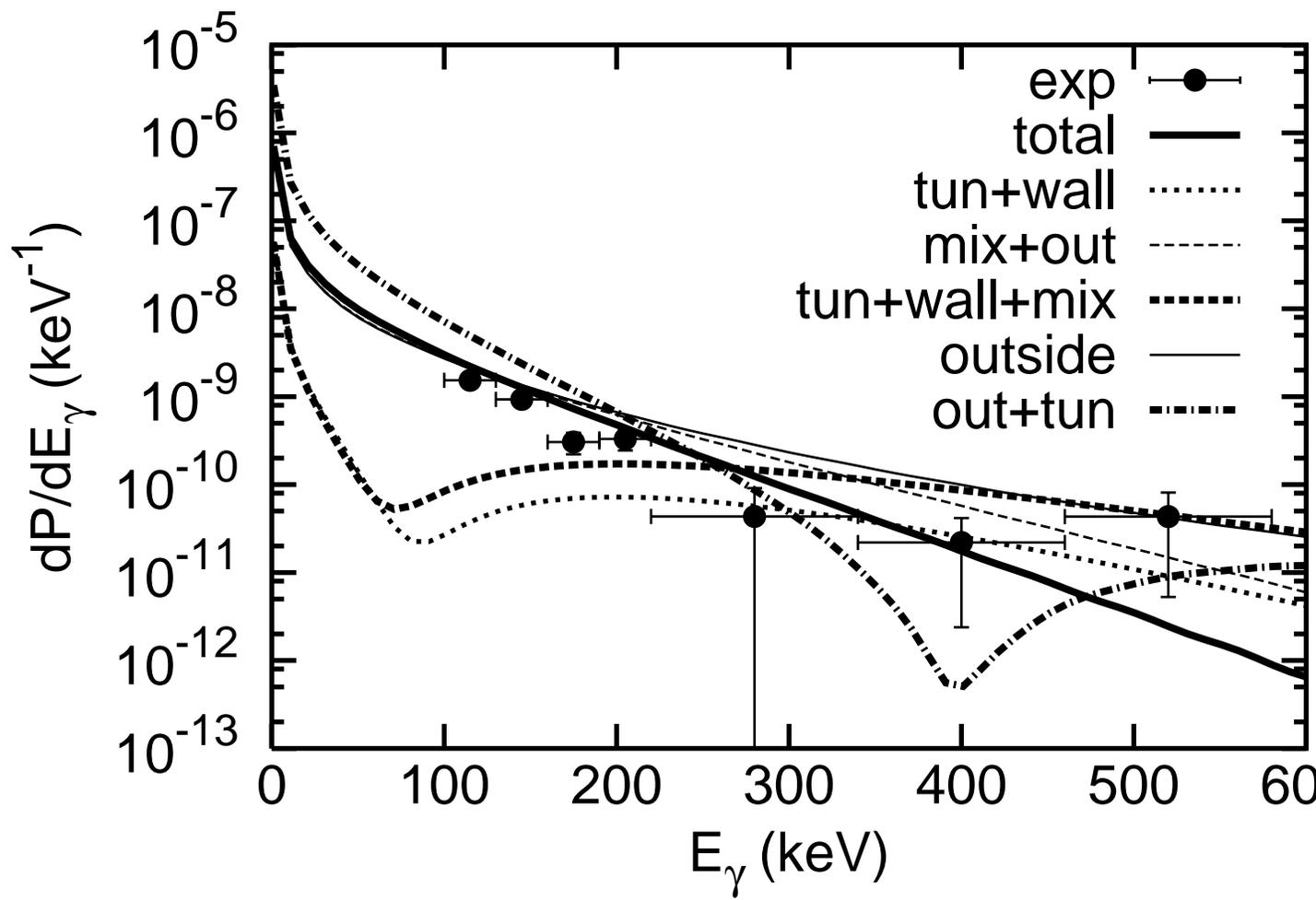
\end{document}